\documentclass[conference]{IEEEtran}

\usepackage{cite}
\usepackage{amsmath,amssymb,amsfonts}
\usepackage{algorithmic}
\usepackage{algorithm}
\usepackage{graphicx}
\usepackage{textcomp}
\usepackage{xcolor}
\usepackage{url}
\usepackage{balance}
\usepackage{verbatim}
\usepackage{booktabs}

\newtheorem{definition}{Definition}

\usepackage{booktabs}
\usepackage{multirow}
\usepackage{adjustbox}

\begin{document}

\title{Efficient Discovery of Conditional Dependencies with Desbordante}
\date{}


\author{
\IEEEauthorblockN{Ivan Kozhukov, Dmitry Fedoseev, Maksim Emelyanov, Artem Smola,\\Pyotr Senichenkov, Pavel Anosov, George Chernishev}
\IEEEauthorblockA{Saint-Petersburg State University \\ 
Saint-Petersburg, Russia\\  \{ivan.s.kozhukov, dmitrii.a.fedoseev, m.emelyanov.m, artem.m.smola, p.senichenkov, pavel.i.anosov, chernishev\}@gmail.com\\}}

\maketitle

\begin{abstract}
Conditional functional dependencies (CFDs) are functional dependencies with a restricted scope: they specify the context in which a dependency holds and are useful for data‑quality tasks, specifying complex integrity constraints, and extracting valuable insights from data.

We study the CFD discovery problem, which is computationally demanding. We build on the state‑of‑the‑art CFDFinder algorithm and introduce a set of algorithmic and engineering improvements, including a parallelization strategy, to produce ParCFDFinder. Our implementation is integrated into Desbordante~--- a high‑performance open‑source data profiler written in C++ that exposes a Python interface, enabling CFD discovery to be invoked from any Python program.

Experimental results show that our enhancements speed up the algorithm by up to $318\times$ ($118\times$ on average) and reduce memory usage by up to $23\times$ ($14\times$ on average) compared with the existing Java-based implementation of Metanome. Integrating ParCFDFinder into Desbordante makes it possible, for the first time, to conveniently discover CFDs on datasets with hundreds of thousands of rows on a commodity machine within a reasonable time.
\end{abstract}

\section{Introduction}

Conditional functional dependencies (CFDs) are a special type of functional dependency~\cite{DBLP:journals/VLDB/PapenbrockEMNRSZN15} that apply not to the entire table but only to a subset of it. 

For example, consider Table~\ref{tbl:example}, which contains product sales data. The CFD
$Product \longrightarrow Price$ for $Country \in \{US, CA\}$ holds in this table, while the functional dependency
$Product \longrightarrow Price$ does not.

\begin{table}[ht] \centering \caption{Sales (example)} \label{tbl:example} \begin{tabular}{rl l r l} \toprule SaleID & Product & Category & Price & Country \\
\midrule 1 & Smartphone X & Electronics & 699.0 & US \\
2 & T-Shirt Classic & Clothing & 25.0 & US \\
3 & Smartphone X & Electronics & 699.0 & CA \\
4 & Office Chair & Furniture & 150.0 & US \\
5 & Smartphone X & Electronics & 749.0 & GB \\
6 & Coffee Maker & Home & 89.0 & GB \\
7 & T-Shirt Classic & Clothing & 25.0 & CA \\
8 & Office Chair & Furniture & 150.0 & CA \\
9 & T-Shirt Classic & Clothing & 23.0 & GB \\
10 & Smartphone X & Electronics & 699.0 & GB \\
\bottomrule \end{tabular} \end{table}

In this context, the problem of discovering CFDs from data arises. Unlike approximate FDs, CFDs provide an explicit context that specifies where a dependency holds, making their scope transparent and enabling discovery of nontrivial, context‑specific insights about the data.

In practice, CFDs enable both knowledge extraction~--- supporting hypothesis generation for scientific discovery~--- and a range of data‑quality interventions (e.g., typo detection, near‑duplicate removal, and schema matching). CFDs also serve to specify complex integrity constraints in data. Finally, in machine‑learning workflows CFDs aid feature engineering and can guide the design of ablation studies.

Desbordante (Spanish for \textit{boundless})~\cite{10.1145/3703323.3703725} is an \textit{open-source} \textit{high-performance} data profiling tool designed for discovery and validation of complex patterns in data. It provides a range of algorithms, including specialized methods for CFD discovery. Tasks in this domain are computationally intensive, and CFD discovery is particularly demanding: it requires both substantial CPU time and large amounts of memory; therefore, algorithms are frequently memory‑bound. Thus, there is a clear need for fast, memory‑efficient discovery algorithms. Currently, there are roughly half a dozen published approaches; the most recent and fastest among them is CFDFinder~\cite{cfdfinder}, which is integrated into the Metanome data profiler~\cite{10.14778/2824032.2824086}.

However, Metanome is more of a research prototype than a production-ready tool. First, it is implemented in Java~--- a language that favors development speed over raw execution speed. As our group's work has shown~\cite{10749955,10143047,10516381,9435469}, simply reimplementing pattern‑discovery algorithms in C++ can yield $2-3\times$ speedups in some (though not all) cases. Second, Metanome lacks a mature user interface, which makes deployment and everyday use difficult. Finally, the Metanome project is no longer actively maintained.

We therefore decided to implement the algorithm in Desbordante and create its high‑performance variant. To that end, we first developed a baseline version and then proposed a set of algorithmic and engineering optimizations, including a parallelization strategy that was absent in the original CFDFinder. The discussion of the design and evaluation of these enhancements forms the core of this paper.

Experimental results on several datasets show that our implementation is faster by a factor of $17-318\times$ (on average $118\times$) and uses $2-23\times$ (on average $14\times$) less memory compared to Metanome.

Overall, we have extended the practical applicability of CFD discovery: for the first time, users can process hundreds of thousands of records on a commodity machine within a reasonable time.

The contributions of this paper are the following:

\begin{itemize}
    \item A set of algorithmic and engineering optimizations for the original CFDFinder algorithm, including a parallelization approach. We call the resulting algorithm ParCFDFinder.
    \item An open‑source implementation of the ParCFDFinder algorithm is included in the Desbordante profiler. Thus, the most efficient CFD discovery algorithm is now ready for practical use and can be employed in any Python program.
    
    \item An experimental evaluation of the proposed algorithm was conducted on multiple datasets, with comparisons against Metanome. We measured both row- and column‑scalability and examined the effect of multi-threading.

    \item A case study that illustrates the interpretation of CFD discovery results using real data.
\end{itemize}

This paper is organized as follows. In Section~\ref{sec:background}, we provide the core definitions necessary for understanding the content of the paper. In Section~\ref{sec:relwork}, we discuss the existing studies concerning CFD discovery. The original CFDFinder algorithm is described in Section~\ref{sec:cfdfinder}, while ParCFDFinder is discussed in Section~\ref{sec:improvements}. The evaluation is presented in Section~\ref{sec:eval}. The case study is detailed in Section~\ref{sec:case-study}. We conclude the paper in Section~\ref{sec:conclusion}.

\section{Background}\label{sec:background}

Let $r$ be a relation over the schema $R$, defined on a set of attributes $\mathcal{A}$.

\begin{definition}
A \textbf{Functional Dependency (FD)} $f: X \rightarrow A$, where $X \subseteq R$ and $A \in R$, \textbf{holds in} $r$ if for all tuples $t_i, t_j \in r$:
$$
t_i[X] = t_j[X] \Longrightarrow t_i[A] = t_j[A].
$$
Here, $X$ is called the \textbf{left-hand side (LHS)} and $A$ the \textbf{right-hand side (RHS)}.
\end{definition}

\begin{definition}
An FD $f: X \rightarrow A$ is a \textbf{generalization} of another FD $g: Y \rightarrow A$ if $X \subset Y$. Conversely, $f$ is a \textbf{specialization} of $g$ if $Y \subset X$.
\end{definition}

\begin{definition}
An FD $f: X \rightarrow A$ is \textbf{minimal} if there exists no attribute $B \in X$ such that the FD $X \setminus B \rightarrow A$ holds in $r$. In other words, $f$ has no generalization that is also valid in $r$.
\end{definition}

\begin{definition}
A \textbf{Conditional Functional Dependency (CFD)} $\varphi$ is a pair $(f: X \rightarrow Y, T_p)$, where $f: X \rightarrow Y$ is a functional dependency (called the \textbf{embedded FD} of $\varphi$) and $T_p$ is a table over attributes $X \cup Y$ (called the \textbf{pattern tableau}). For every attribute $A \in X \cup Y$ and every tuple $t_p \in T_p$, the value $t_p[A]$ is either a constant $a \in \text{dom}(A)$ or the wildcard symbol ``\_'' (which matches any value from $\text{dom}(A)$). A tuple $t_p \in T_p$ is called a \textbf{pattern tuple} (or simply a \textbf{pattern}).
\end{definition}

\begin{definition}
A tuple $t \in r$ \textbf{matches} a pattern $t_p \in T_p$ on an attribute set $S \subset R$, denoted $t \asymp t_p$, if for every attribute $B \in S$: $t_p[B] =$ ``\_'' or $t_p[B] = t[B]$.
\end{definition}

\begin{definition}
A relation $r$ \textbf{satisfies a CFD} $\varphi = (X \rightarrow Y, T_p)$ if for every $t_i, t_j \in r$ and every pattern $t_p \in T_p$:
\[
t_i[X] = t_j[X] \asymp t_p[X] \Longrightarrow t_i[Y] = t_j[Y] \asymp t_p[Y].
\]
\end{definition}

\begin{definition}
For a CFD $\varphi = (X \rightarrow A, T_p)$ and a relation $r$, the \textbf{cover} of a pattern $p \in T_p$ is 
defined as:
\[
\text{cover}(p) = \{ t \in r \mid t[X] \asymp p[X] \}.
\]
\end{definition}

\begin{definition}
The \textbf{local support} of a pattern $p$ is defined as:
\[
\text{local\_support}(p) = \frac{|\text{cover}(p)|}{|r|}.
\]
\end{definition}

\begin{definition}
The \textbf{global support} of a pattern tableau $T_p$ is defined as:
\[
\text{global\_support}(T_p) = \frac{\left| \bigcup_{p \in T_p} \text{cover}(p) \right|}{|r|}.
\]
\end{definition}

We refer to the \textbf{support of a CFD} $\varphi$ as the global support of its pattern tableau.

\begin{definition}
The set of \textbf{keepers} of a pattern $p$ is the set of tuples covered by $p$ that do not cause a violation of the embedded FD or any single-tuple violation.
\end{definition}

\begin{definition}
The \textbf{local confidence} of a pattern $p$ is defined as:
\[
\text{local\_confidence}(p) = \frac{|\text{keepers}(p)|}{|\text{cover}(p)|}.
\]
\end{definition}

\begin{definition}
The \textbf{global confidence} of a pattern tableau $T_p$ is defined as:
\[
\text{global\_confidence}(T_p) = \frac{\left| \bigcup_{p \in T_p} \text{keepers}(p) \right|}{\left| \bigcup_{p \in T_p} \text{cover}(p) \right|}.
\]
\end{definition}

We refer to the \textbf{confidence of a CFD} $\varphi$ as the global confidence of its pattern tableau.

\section{Related Work}\label{sec:relwork}
\subsection{Discovering data quality rules}
In their seminal work, Chiang and Miller~\cite{DiscDataQualRules} introduced several metrics~--- such as \textit{Support}, $\chi^2$, and 
\textit{Conviction}~--- to identify interesting Conditional Functional Dependencies (CFDs). These metrics help filter out trivial dependencies, thus preventing the common issue where discovery algorithms generate an overwhelming number of CFDs, only a few of which are truly significant. While this aspect aligns with many studies in CFD discovery, a distinctive feature of their work is an algorithm that performs two key tasks: discovering CFDs and detecting violations of them. These violations can then be flagged for data analysts to review and correct.

The experimental evaluation revealed two key complexity characteristics of the algorithm: it scales exponentially with the number of attributes and approximately linearly with the domain size of the attributes. Furthermore, the study evaluated and compared several metrics for quantifying the ``interestingness'' of discovered CFDs. The results demonstrated that \textit{Conviction} outperformed the others, proving to be the superior measure for highlighting the most useful dependencies. The metrics of \textit{Confidence}, \textit{Support}, and \textit{Interest} followed in decreasing order of effectiveness, though they still provided valuable results. 

Importantly, the authors investigated the use of \textit{Support} and \textit{Conviction} metrics for identifying data exceptions (i.e., CFD violations). Detecting exceptions via \textit{Support} highlighted infrequently occurring values, which are not necessarily errors. In contrast, using \textit{Conviction} targeted data that violate an approximate CFD due to statistical independence. Their experimental results demonstrated an advantage of \textit{Conviction}: it yielded a much richer set of interesting violations for analyst review.

\subsection{Discovering Conditional Functional Dependencies}

Fan et al.~\cite{FanCFD} present three CFD mining algorithms that differ in their features and applicability: CFDMiner, CTANE, and FastCFD.

\subsubsection{CFDMiner}
CFDMiner is an efficient algorithm for discovering constant CFDs, namely minimal, k-frequent, and left-reduced constant CFDs. A key feature of the algorithm is that the authors successfully reduce the problem of finding minimal constant CFDs to the established problem of discovering all k-frequent closed itemsets~\cite{closed-itemsets}. Afterwards, the authors matched them with their corresponding free itemsets.

\subsubsection{CTANE}
CTANE is an extension of the TANE algorithm for levelwise discovery of FDs. The output of CTANE is the set of all minimal, k-frequent CFDs of a relational instance, i.e., its complete canonical cover. The principle of CTANE's operation is based on a systematic, levelwise enumeration of attribute combinations $X$ and patterns $t_p$ (patterns may contain wildcards). The algorithm starts with single attributes and at each subsequent level constructs more complex candidates. Its efficiency is achieved through a specialized mechanism of auxiliary structures $C^+$, which precompute possible right-hand sides for each candidate and prune provably unpromising search branches.

\subsubsection{FastCFD}
The FastCFD algorithm, like CTANE, discovers all minimal k-frequent CFDs; however, it employs a depth-first search strategy instead of a levelwise traversal. The set of optimizations used in the algorithm, such as decomposing the problem into subproblems for each RHS and utilizing minimal covers of difference sets, is primarily aimed at handling data with high arity (more than 15 attributes). Furthermore, FastCFD leverages the results produced by CFDMiner in its operation.

\subsubsection{Applicability Scope}
The experimental results provided by the authors suggest potential application scenarios for these algorithms. CFDMiner can be employed for discovering exclusively constant CFDs. CTANE is applicable to datasets with low arity and high support thresholds, while FastCFD is suitable for scenarios featuring high arity and moderate data volumes.

\subsection{Discovering (frequent) Constant Conditional Functional Dependencies}
Diallo, Novelli, and Petit~\cite{DialloCFUN} proposed the CFUN algorithm, which, similar to CFDMiner, focuses exclusively on discovering frequent \textit{constant} CFDs. However, unlike the approaches based on itemsets (like CFDMiner) or hypergraphs, CFUN is an extension of the FUN algorithm (originally designed for standard functional dependencies) that has been adapted for the conditional setting. The algorithm introduces the concepts of \textit{conditional agree sets}, \textit{conditional closure}, and \textit{quasi-closure}. Operating in a levelwise manner (akin to Apriori), CFUN identifies the canonical cover of satisfied CFDs by computing the difference between the closure and quasi-closure of candidate attribute sets.

Regarding applicability, the authors demonstrate that CFUN scales linearly with the number of tuples, making it feasible for datasets with a large number of rows. Its memory usage is also shown to be efficient. Unlike generalized algorithms (like CTANE), CFUN is optimized specifically for constant patterns, offering a direct method to derive a non-redundant cover without the intermediate step of mining closed itemsets.

\subsection{Splitting the ``C'' from the ``FD''}
Rammelaere and Geerts~\cite{Rammelaere2019} revisited the problem of approximate CFD discovery, presenting it as a combination of functional dependency (FD) discovery and pattern mining. They described three approaches that differ in how they traverse the search space:
\subsubsection{Integrated Approach} This strategy traverses a combined lattice of constant and variable patterns simultaneously. This is the approach employed by the classic CTANE algorithm.

\subsubsection{Itemset-First} This new approach decouples the process by first performing frequent itemset mining to discover constant patterns (the ``C''). Then, for each pattern found, a data projection is formed, on which the functional dependency discovery algorithm is run.

\subsubsection{FD-First} This strategy summarizes the idea behind the work \cite{DiscDataQualRules} by initially searching for variable FDs on the entire dataset. For FDs that do not hold globally but have sufficient confidence, it subsequently runs a specialized pattern mining step to identify the specific constant conditions under which these dependencies are valid.

\begin{figure}
    \small
    \centering
    \includegraphics[width=0.3\textwidth]{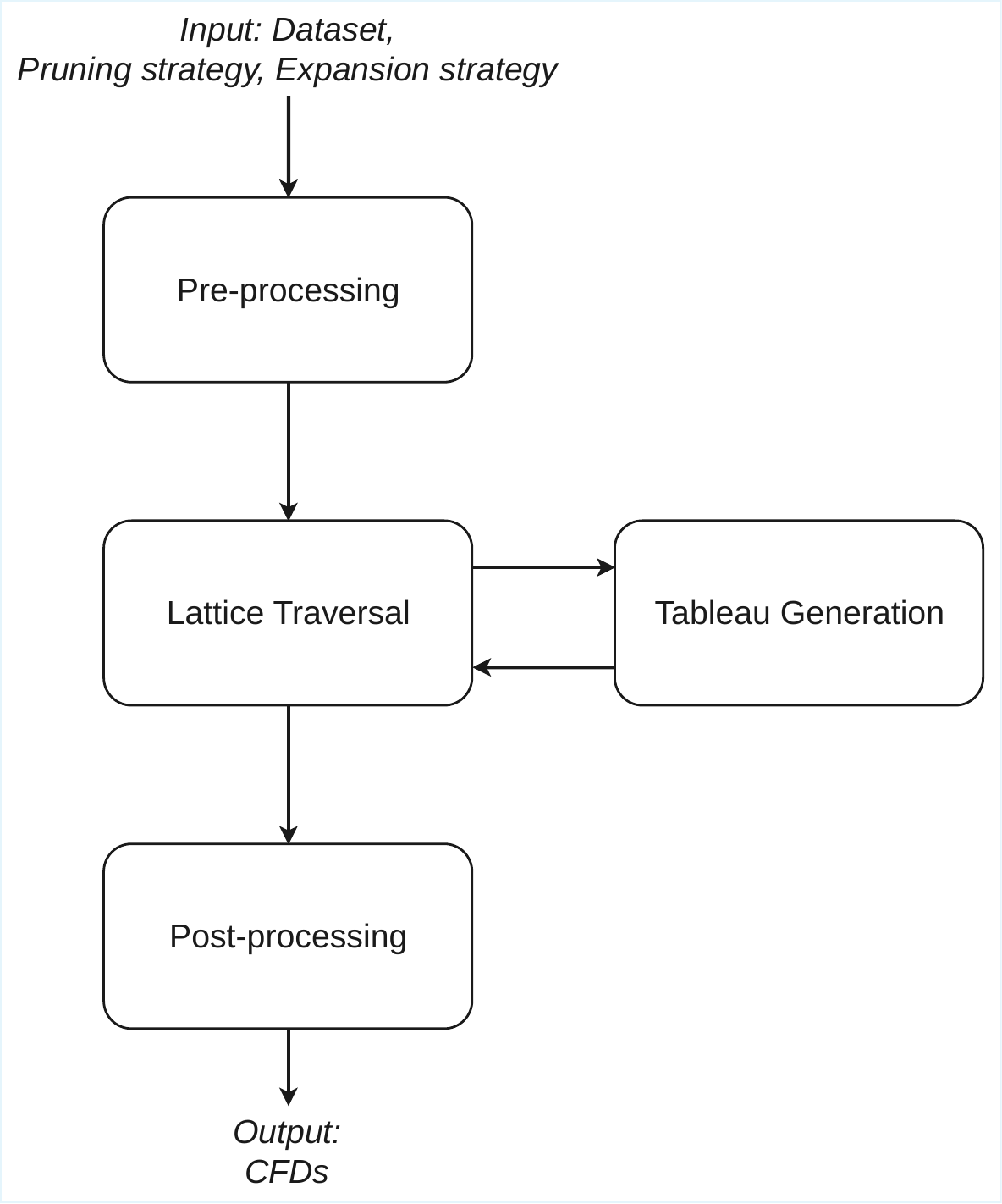}
    \caption{Schema of CFDFinder algorithm}
    \label{Schema}
\end{figure}

\newpage
\section{CFDFinder}\label{sec:cfdfinder}

This section provides a step-by-step analysis of the CFDFinder algorithm, aimed at a deeper understanding of the approach under consideration. All stages of the algorithm are schematically shown in Figure~\ref{Schema}. In addition, some implementation details that are key to the optimization methods we propose are discussed here.

\subsection{Pre-processing}

During the pre-processing stage, CFDFinder constructs auxiliary structures and identifies the complete set of maximal non-FDs for the relation. This set subsequently serves as the foundation for generating CFD candidates. At this stage, CFDFinder employs a modified version of the HyFD FD discovery algorithm \cite{HyFD}.

The initial stage of HyFD involves creating \textbf{Position List Indexes (PLIs)} and \textbf{compressed records}. A PLI is constructed for a specific attribute $X$ as a collection of index lists, where each list groups the row indices (row IDs) that share the same value for attribute $X$. Effectively, a PLI partitions the relation into a set of equivalence classes, referred to as \textbf{clusters}. The compressed representation of records is then generated by replacing each attribute value with the index of the cluster to which the corresponding record belongs. 

Subsequently, the operation of the HyFD algorithm consists of three repeating stages: \textbf{sampling}, \textbf{induction}, and \textbf{validation}. During the \textbf{sampling stage}, pairs of records are selected that match only on some attributes and have different values for the remaining attributes. Based on these pairs, a set of non-FDs of the form $X \rightarrow A$ is constructed, where $X$ is the set of matching attributes and $A$ is the mismatching attribute. The task of iterating through all possible pairs of tuples has quadratic complexity, which results in poor scalability. To avoid performance issues, only a subset of record pairs is selected via specialized sampling techniques during the sampling stage. At the \textbf{induction stage}, FD candidates are formed based on the set of non-FDs obtained during the sampling stage. These candidates are then verified during the \textbf{validation stage}. During both the induction and validation stages, non-FDs are identified, including those dependencies that do not become candidates during induction and those that become candidates but were determined to be non-FDs during validation. In the original HyFD, these non-FDs are discarded, but in CFDFinder, they are added to a set of non-FDs from which any non-FD with a specialization in the tree is removed, meaning only maximal non-FDs remain.

The final output of HyFD is a set of minimal FDs. CFDFinder augments the accumulated maximal non-FDs by adding dependencies derived from these minimal FDs~--- those obtained by removing a single attribute from the left-hand side. Only those generalizations that are themselves maximal non-FDs (i.e., those for which no specialization exists in the current set) are included. The obtained set of maximal non-FDs is the set of all maximal non-FDs of the relation.

\subsection{Lattice traversal}

During the lattice traversal stage, an attempt is made to generate a pattern tableau for CFD candidates, and new candidates are created based on existing ones. The pseudocode for this part of the algorithm is presented in Algorithm \ref{alg:lattice-traversal}. 

\begin{algorithm}
\small
\caption{Top-down lattice traversal}
\label{alg:lattice-traversal}
\begin{algorithmic}[1]
\STATE \textbf{Data:} $N$ --- set of maximal non-FDs, $R$ --- relation
\FOR{$1 \leq i \leq |R|$}
    \STATE $L_i \gets \emptyset$
\ENDFOR
\FOR{$n = (X_n, A_n) \in N$}
    \STATE $l = |X_n|$
    \STATE $L_l \gets L_l \cup \{n\}$
\ENDFOR
\STATE $p \gets |R|$
\WHILE{$p > 0$}
    \FOR{$c = (X_c, A_c) \in L_p$}
        \STATE $T \gets buildTableau(c)$
        \IF{$T \neq \emptyset$}
            \STATE $S \gets \{(X',A_c) \mid X' \subset X_c \land |X_c \setminus X'| = 1\}$
            \STATE $L_{p-1} \gets L_{p-1} \cup S$
        \ENDIF
    \ENDFOR
    \STATE $p \gets p-1$
\ENDWHILE
\end{algorithmic}
\end{algorithm}

First, all remaining levels of the lattice are prepared (lines 2--4 of the pseudocode). Then, initial candidates from the set of maximal non-FDs obtained in the previous step are added to their corresponding levels; the level number corresponds to the number of attributes in the candidate’s LHS (lines 5--8).

Next, the lattice is traversed top-down, from the level of candidates with the maximum number of attributes in the relation down to the level of candidates with a single attribute (lines 9--19). For each candidate, an attempt is made to create a pattern tableau, the validity of which is determined according to the pruning strategy. If a valid tableau is obtained for a candidate, that CFD candidate is considered a valid CFD, and new candidates~--- generated by removing one attribute from the LHS of the resulting CFD~--- are added to the lattice at the level below.

\subsection{Pattern Tableau generation}

In the pattern tableau generation stage, the algorithm attempts to construct a pattern tableau for a CFD candidate. It iteratively adds new patterns to the tableau as long as the chosen strategy permits. The stage concludes when no further patterns can be added. The resulting tableau is then evaluated against predefined mining parameters (e.g., confidence and support thresholds). If it satisfies these constraints, the corresponding CFD is recorded in the final set of mining results. The pattern tableau generation stage is presented in Algorithm \ref{alg:pattern-tableau}.

\begin{algorithm}
\small
\caption{The tableau generation algorithm of CFDFinder}
\label{alg:pattern-tableau}
\small
\begin{algorithmic}[1]
\STATE \textbf{Data:} d~--- CFD candidate, r~--- relation instance, ps~--- pruning strategy, es~--- expansion strategy 
\STATE $T\leftarrow \emptyset; F \leftarrow \emptyset; p_0 \leftarrow es.generateNullPattern(d)$
\FOR{\textbf{all} $t\in r$}
    \STATE $cover(p_0) \leftarrow cover(p_0)\cup \{t\}$
    \STATE $kp(p_0) \leftarrow kp(p_0) + kp(t)$
    \STATE $ct(p_0) \leftarrow ct(p_0) + ct(t)$
\ENDFOR
\STATE $margSupport(p_0) \leftarrow ct(p_0)$
\STATE $conf(p_0) \leftarrow kp(p_0)/ct(p_0)$
\STATE $F \leftarrow F  \cup \{p_0\}$
\WHILE{$F\neq \emptyset$ and $ps.continue(T)$}
\STATE $p \leftarrow argmax_{x\in F} margSupp(x)$
\STATE $F \leftarrow F \setminus \{p \}$
\IF{$ps.addPattern(p)$}
\STATE $T \leftarrow T \cup \{p \}$
\FOR{\textbf{all} $p' \in F$}
\STATE $cover(p') \leftarrow cover(p') \setminus cover(p)$
\STATE $margSupport(p') \leftarrow \Sigma_{t\in cover(p')}ct(t)$
\IF{$\neg ps.considerPattern(p')$}
\STATE $F \leftarrow F \setminus \{ p' \}$
\ENDIF
\ENDFOR
\ELSE
\FOR{\textbf{all} $c \in es.expand(p)$}
\IF{$ps.valid(c)$}
\STATE $cover(c) \leftarrow determineCover(cover(p), c)$
\STATE $margSupport(c) \leftarrow |cover(c)|$
\IF{$ps.considerPattern(c)$}
\STATE $F \leftarrow F \cup \{ c \}$
\ENDIF
\ENDIF
\ENDFOR
\ENDIF
\ENDWHILE
\IF {$ps.continueGeneration(T)$}
\STATE $return~T$
\ENDIF
\STATE $return~\emptyset$
\end{algorithmic}
\end{algorithm}

Before delving into the details of the algorithm, we introduce a few additional definitions.

\begin{definition}
A pattern $p_p$ that contains exactly one more specified constant than $p_c$ is called a \textbf{parent pattern} of $p_c$. Similarly, a pattern $p_c$ that contains exactly one fewer specified constant than $p_p$ is called a \textbf{child pattern} of $p_p$. A single pattern may have multiple child patterns and multiple parent patterns.
\end{definition}

\begin{definition}
A pattern that consists exclusively of the wildcard symbol ``\_'' for all attributes is called the \textbf{null pattern}.
\end{definition}

The tableau generation algorithm is heuristic. It relies on the premise that if the selected patterns individually exhibit a local confidence higher than a predefined global confidence threshold, then the entire tableau will also meet or exceed that threshold.

Initially, the algorithm creates a \textbf{null pattern} \( p_0 \) and adds all unique tuples to its cover. This is possible because the null pattern matches every tuple. In the listing, \( kp(p) \) denotes the number of keepers of pattern \( p \), and \( ct(p) \) represents the size of its cover (i.e., \( |\text{cover}(p)| \)).

The cover, support, and confidence of the null pattern are then computed. Afterwards, \( p_0 \) is added to the \textbf{frontier} \( F \). The frontier is a set of pattern candidates sorted by \textbf{marginal support}~--- the additional support a pattern would contribute to the current tableau. This ordering enables a greedy strategy that always selects the candidate with the highest marginal support.

Next, the algorithm checks whether the current tableau \( T \) can be improved according to the chosen strategy via the \( \mathit{continue}(T) \) function. For instance, this function may verify whether the tableau already meets a fixed support threshold. If the check passes, the algorithm proceeds to process the pattern with the highest marginal support from the frontier.

If a pattern \( p \) can be added to the tableau, its cover is subtracted from the cover of every other candidate \( p' \) in the frontier. This update ensures that the \textbf{marginal support} of each candidate correctly reflects only the tuples not yet covered by \( T \), thereby minimizing overlap and guaranteeing that the global support increases by exactly the marginal support of \( p \). Additionally, the function $considerPattern(p')$ determines whether it is still worthwhile to keep \( p' \) in the frontier. For example, candidates with zero marginal support are removed to improve performance.

If a pattern $p$ is \textbf{not} selected for inclusion, it is \textbf{expanded}: all its child patterns are generated. The on-demand expansion~\cite{Expansions} uses the $\mathit{valid}(c)$ function. According to the selected pruning strategy, this function verifies either the completion of processing for all parent nodes or the absence of the current pattern from the set of previously visited ones. If $c$ satisfies the mining parameters, it is added to the frontier. The cover of a child pattern can be computed efficiently from the cover of its parent with Algorithm~\ref{child_cover}, reducing the computational overhead of repeatedly scanning the entire relation.

\begin{algorithm}
\small
\caption{Computing the cover of a child pattern}
\label{child_cover}
\small
\begin{algorithmic}[1]
\STATE \textbf{Data:} $p_c$~--- child pattern, $cover(p)$~--- parent cover
\STATE $cover_c \leftarrow \emptyset$
\FOR {\textbf{each} $cluster \in cover(p)$}
    \STATE $matches \leftarrow true$
    \STATE $t \leftarrow cluster[0]$
    \FOR {$A \in attr(p_c)$}
    \IF{$\neg p_c[A].Matches(t[A])$} 
        \STATE $matches \leftarrow false$
    \ENDIF
    \ENDFOR

    \IF{$matches = true$}
        \STATE $cover_c \leftarrow cover_c \cup cluster$
    \ENDIF
\ENDFOR
\STATE \textbf{return} $cover_c$
\end{algorithmic}
\end{algorithm}

The algorithm terminates when the frontier is empty or when a stopping condition defined by the strategy is met. It then either returns a valid pattern tableau for the CFD candidate or~--- if the strategy’s pruning criteria cannot be satisfied~--- returns an empty set.

\subsection{Pruning Strategies}
CFDFinder implements several pruning strategies.

\subsubsection{Minimum Confidence and Support Thresholds}
According to this pruning strategy, tableau generation continues if the frontier contains at least one pattern and if the minimum support of the pattern tableau has not reached a certain specified value. Only patterns with local confidence no less than the specified value are added to the tableau, while patterns with zero support are discarded. A CFD candidate is added to the resulting CFD set if the generated pattern tableau satisfies the specified thresholds.

\subsubsection{Minimum Support Gain and Maximum Support Drop Thresholds (Support Independent Strategy)}
\begin{definition}
\textbf{Minimum support gain} is the lower bound of the local support of patterns in the tableau.
\end{definition}

\begin{definition}
\textbf{Maximum support drop} is the upper bound of the difference between the support of the current CFD candidate and the support of its generalizations.
\end{definition}

\begin{definition}
\textbf{Maximum pattern tableau length} is the upper limit of the number of patterns in the pattern tableau.
\end{definition}

The algorithm terminates its operation if any of the following conditions is met: 1) there are no patterns remaining in the frontier, or 2) the number of patterns in the tableau equals the maximum tableau length, or 3) there are no patterns in the frontier that meet the minimum support gain threshold.

Using this strategy, we consider only those patterns whose support gain is greater than the specified threshold. A CFD candidate is added to the resulting set if the generated pattern tableau satisfies the specified confidence threshold and the support drop does not exceed the specified upper bound.

\subsubsection{Partial FD}

\begin{definition}
The $\textbf{g1}$ metric is the fraction of record pairs violating a given FD.
\end{definition}

\begin{definition}
Let $s \in [0,1]$; then a \textbf{partial FD} $f: X \rightarrow A$, where $X \subseteq R$ and $A \in R$, is said to \textbf{hold in} $r$ if $g1 \leq s$.
\end{definition}

When this strategy is chosen, no patterns are generated except for null patterns. The tableau creation function \textit{buildTableau} is simplified to generating only the null pattern and returning a tableau containing only the null pattern if the $g1$ metric value for this pattern does not exceed a specified threshold value; otherwise, an empty tableau is returned.

It is easy to see that the result of the algorithm's operation with the $g1$ strategy is a set of partial FDs.

\subsection{Expansion Strategies}

During the expansion stage, the algorithm processes patterns that have sufficient local support but lack the required local confidence to be included directly in the resulting pattern tableau. These patterns cannot be pruned entirely, as they may lead to viable specializations. The idea is to specialize them into new pattern candidates and add these candidates to the frontier. A key property of specialization is that the cover of any child pattern is always a subset of the cover of its parent, regardless of the specific expansion strategy employed.

An \textbf{expansion strategy} defines a method for enumerating child patterns. The CFDFinder algorithm employs three expansion strategies proposed in~\cite{Expansions}. 

\subsubsection{Expansion by Adding Constants}

This strategy is suitable for standard CFDs, where attributes may take either constant values or the wildcard symbol ``\_''. A pattern is specialized by replacing each wildcard symbol with constants drawn from the attribute's domain. This approach generates all possible combinations of constants across the wildcard positions, thereby producing a complete set of child patterns. The process is illustrated in Algorithm~\ref{Constant_expansion}.

\begin{algorithm}
\small
\caption{Expansion by Adding Constants}
\label{Constant_expansion}
\begin{algorithmic}[1]
\STATE $children \leftarrow \emptyset$
\FOR {$t \in cover(p)$}
    \FOR {$A \in attr(p)$}
    \IF{$p[A] = $``\_''} 
        \STATE $c \leftarrow p$
        \STATE $c[A] \leftarrow t[A]$
        \STATE $children \leftarrow children \cup \{c\}$
    \ENDIF
    \ENDFOR
\ENDFOR
\STATE \textbf{return} $children$
\end{algorithmic}
\end{algorithm}

\subsubsection{Expansion for Negative Conditions}

To support negative conditions, the assignment operation $c[A] \leftarrow t[A]$ must be extended. For each constant $a \in \text{dom}(A)$, we now generate two types of conditions: equality $A = a$ and inequality $A \neq a$. This modification doubles the search space and significantly affects the pruning process.

Specializing $p$ by constraining an attribute $A$ to a value $a$ yields two child patterns: $c_{[=a]}$ (with $A = a$) and $c_{[\neq a]}$ (with $A \neq a$). Their covers partition $cover(p)$: the tuples covered by $c_{[\neq a]}$ are exactly those in $cover(p)$ not covered by $c_{[=a]}$.

\subsubsection{Expansion for Range Conditions}

This strategy employs a range-based representation of attribute bindings. We assume a total order on each attribute domain and assign indices to values accordingly. For a CFD $\varphi: (X \rightarrow Y, T_p)$, each pattern tuple $t_p \in T_p$ assigns a range $t_p[A] = [l, r]$ to every attribute $A \in X \cup Y$, where $l$ and $r$ are indices of values in $\text{dom}(A)$ and $l \leq r$. A data tuple $t$ satisfies this condition iff the value $t[A]$ corresponds to an index $i$ such that $l \leq i \leq r$. Range bindings generalize both wildcard symbols (when $l$ is the first index and $r$ is the last) and constant bindings (when $l = r$ corresponds to a single value).

The null pattern $p_0$ is initialized with ranges covering the entire domain of each attribute (i.e., from the first index to the last). During expansion, for each attribute $A$ with $p[A] = [l, r]$ and $l < r$, two child patterns are generated: $[l+1, r]$ and $[l, r-1]$. This effectively enumerates all contiguous subranges by removing one boundary index at a time.

\section{Proposed Approach}\label{sec:improvements}

The Java version of the CFDFinder algorithm, developed within the Metanome data profiling framework\footnote{\url{http://www.metanome.de}}, was taken as the baseline version. This implementation was originally described in~\cite{cfdfinder}.

We implemented two improved variants of the algorithm in C++ within the high-performance data profiler Desbordante. The first version is a re-implementation of the Java version of the algorithm with minimal changes necessitated by differences in the data structure implementations in Java and C++. The second version contains a series of our improvements and code optimizations that provide a significant increase in the performance and scalability of the algorithm.

\subsection{C++ Re-implementation}
In the Java version, the frontier is implemented using \textit{PriorityQueue}. When porting to C++, the equivalent \textit{std::priority\_queue} from the STL was considered. However, it lacks a method to search for elements within the queue, which is required for the algorithm. Therefore, it was replaced with \textit{boost::multi\_index}, which allows for the definition of two indices simultaneously: one for priority ordering (similar to a priority queue) and one for hash-based element lookup. 

Overall, the C++ port did not yield significant performance improvement on most datasets. However, for datasets where the Java version spent a substantial portion of its execution time on frontier lookups, the replacement with \textit{boost::multi\_index} resulted in a noticeable speedup. This observation motivated us to pursue further algorithmic and technical optimizations.

\subsection{CFDFinder Optimization}

\subsubsection{Improved Structures}

After generating a child pattern during the expansion process, the algorithm, according to the selected pruning strategy, checks whether all parent patterns have been fully considered or whether the current pattern has already been processed. The \textbf{visited} structure, representing a set of visited patterns, is used for this check. In the Java version, this is implemented using a \textit{HashSet} (a hash table using the chaining method). In the improved version, we used \textit{boost::unordered\_flat\_set} instead of \textit{std::unordered\_set} from the C++ Standard Template Library. This container provides higher performance due to its use of open addressing and the resulting good data locality.

\subsubsection{Pattern Generation Optimizations}

The pattern generation stage is the most computationally intensive part of the algorithm, accounting for approximately 99\% of the total execution time across all tested datasets. Our optimizations target this stage by leveraging a key observation: a child pattern differs from its parent in only one attribute. This allows us to restructure the workflow: instead of immediately constructing full child patterns, we first work with lightweight \emph{candidates} --- potential patterns represented only by the attribute binding being specialized. All validation and filtering steps are performed using only this binding and the parent's metadata. Only after a candidate passes all checks do we materialize the full pattern and its cover.

We organize the optimizations into four complementary techniques, each addressing a specific inefficiency in the original implementation.

\paragraph{Eliminating Duplicate Pattern Generation}
The Constant and NegativeConstant expansion strategies generate child patterns by replacing a wildcard with values taken from representative tuples of each cluster in the parent's cover. Different clusters often share the same value for the expanded attribute, leading to duplicate child patterns. In the original algorithm, each duplicate undergoes full cover construction and validation before being discarded during the duplicate check.

Our solution preprocesses the cluster representatives to extract only \emph{unique values} for the attribute being expanded. This simple filtering eliminates redundant computations and reduces memory allocations by ensuring that each distinct value generates at most one candidate pattern.

\paragraph{Deferred Cover Construction with Bitmask Pruning}
Originally, to determine whether a child pattern meets the support threshold, the algorithm first constructed its full cover (by copying clusters from the parent cover), computed support, and then discarded the cover if validation failed. This resulted in excessive memory operations.

We introduce a lightweight \textbf{cover mask}~--- a bitset with a length equal to the number of clusters in the parent cover. For each candidate child pattern (identified by its attribute binding), we compute the mask in a single pass over the parent clusters, setting a bit for each matching cluster. From this mask, we can calculate the pattern's support without materializing the cover. Only if the support satisfies the threshold do we proceed to construct the actual cover by selecting the clusters indicated by the mask. This approach eliminates unnecessary copying and reduces memory pressure.

\paragraph{Batch Processing of Multiple Candidates}
The original implementation processed each child pattern sequentially: for each generated candidate, it performed validation, support calculation, and (if successful) cover construction. This pattern-by-pattern approach exhibits poor data locality: each iteration operates on different attribute bindings, causing frequent cache misses and preventing the compiler from applying vectorized optimizations.

We restructure the workflow into a batched pipeline that processes all candidates for a given attribute together. For the attribute being expanded, we precompute the values from each parent cluster representative and store them in a contiguous array. This data layout enables efficient cache utilization and allows the subsequent operations to be performed uniformly across all candidates:

\begin{enumerate}
    \item Generate all candidate child patterns from the parent (using unique values to avoid duplicates).
    \item Apply validation filters to the entire batch, removing invalid candidates early.
    \item Compute cover masks and support for all remaining candidates in a single pass over the precomputed attribute values.
    \item Filter candidates that fail support thresholds.
    \item Finally, construct full covers only for the surviving patterns.
\end{enumerate}

This batch-oriented approach progressively reduces the number of candidates before performing expensive operations, enables uniform processing across candidates, and significantly improves cache locality by operating on contiguously stored data.

\paragraph{Reusing Intermediate Results for Negative Constants}
For the \textit{NegativeConstantStrategy}, after computing the cover mask for a valid constant, the mask for its negation is obtained simply as the bitwise complement. Similarly, its support is computed as the difference between the total number of rows and the support of the constant being negated. Our batched pipeline naturally reuses these intermediate results, thereby avoiding the redundant independent computations performed in the original implementation.

\subsubsection{Parallelization of lattice traversal}

The pattern tableau generation stage for each CFD candidate is computationally expensive, as it involves exploring a large space of candidate patterns. Fortunately, these generation tasks are independent: the attempt to construct a tableau for one candidate does not affect the others. This means that when traversing candidates of the same level, we can process them in parallel, as noted in~\cite{cfdfinder}.

To leverage this property, we parallelized the lattice traversal using a thread pool implemented using the Boost.Asio library. We divide the list of candidates into several batches and process them concurrently using the thread pool. The pool distributes the batches across multiple CPU cores, executing multiple tableau generation tasks in parallel. Boost.Asio provides an efficient work-stealing mechanism that balances the load dynamically, ensuring that cores remain busy even when some tasks complete earlier than others. This approach significantly reduces the overall runtime of the mining process and scales well with the number of available cores.

\section{Evaluation}\label{sec:eval}
\subsection{Methodology}

To verify the correctness of our C++ implementation, we compare its mined CFDs with those produced by the original CFDFinder algorithm integrated into the Metanome framework.

During testing, we observed that the original algorithm does not fully specify the order in which candidate patterns are traversed. Specifically, when multiple patterns have identical support and confidence values, the order of selection becomes ambiguous. Depending on the chosen order, a different pattern may be added to the tableau, which in turn affects the recalculation of candidate parameters, leading to divergent subsequent choices. As a result, the final pattern tableau may differ, making it impossible to determine whether discrepancies stem from implementation errors or merely from non-deterministic traversal. Furthermore, this non-determinism also affects execution time, as different traversal orders lead to processing different sets of candidates and performing different numbers of support recomputations, rendering runtime comparisons equally unreliable.

To address this issue, we introduced additional ordering criteria that enforce a deterministic and unambiguous traversal of candidates. The primary ordering remains based on support and confidence; however, when ties occur, we proceed to compare the attribute values of the patterns. Each pattern is represented as a list of elements, each encoding a condition determined by the chosen expansion strategy. For each type of condition, we define a fixed local order; the exact ordering is not critical as long as it is applied consistently and identically across both implementations. With this deterministic ordering in place, the two implementations can be compared correctly and reproducibly.

Reference outputs are obtained by running the modified version of the Metanome implementation under the same strategies and parameter settings.

\subsection{Experimental Setup and Configuration}

\paragraph{Hardware and software}

Experimental studies were conducted on a testbed with the following hardware and software configurations. Hardware: Intel Xeon Gold 6338 (Icelake), 4 physical cores (8 threads, MT), 48 GiB RAM, 16 MiB L2 Cache. Software: Ubuntu 24.04.4 LTS (Noble Numbat), gcc (Ubuntu 13.3.0-6ubuntu2~24.04.1) 13.3.0, CMake 3.28.3, Boost 1.89.0, openjdk 1.8.0\_482,  OpenJDK Runtime Environment (build 1.8.0\_482-8u482-ga~us1-0ubuntu1~24.04-b08), OpenJDK 64-Bit Server VM (build 25.482-b08, mixed mode).

\paragraph{Datasets}
To conduct the testing, a collection of datasets was assembled, including both real-world and synthetic data. Several of these datasets were used in the paper~\cite{cfdfinder}. These include the following datasets: Wisconsin Breast Cancer, Bridges, Echocardiogram, ncvoter, abalone and Wine Reviews. Additional datasets were added to ensure a more comprehensive study. These include: Students\footnote{URL:\url{https://www.kaggle.com/datasets/amar5693/student-performance-dataset}}, BMW Global Sales\footnote{URL:\url{https://www.kaggle.com/datasets/payaldhokane/bmw-global-sales-and-market-data?resource=download}}, Biocase multimedia object and Biocase gathering namedareas\footnote{URL:\url{https://my.hidrive.com/share/rt.v.6myak#$/}}. The characteristics of the datasets are provided in Table \ref{table_datasets}.

\begin{table}[htbp] \centering \caption{Used 
datasets} \label{table_datasets} \begin{tabular}{l c c c} \toprule Dataset & Num\_rows & Num\_cols & Size (KB) \\
\midrule Bridges & 108 & 13 & 6 \\
Echocardiogram & 132 & 13 & 6 \\
Wisconsin Breast Cancer & 699 & 11 & 20 \\
BMW Global Sales & 1000 & 14 & 69.67 \\
ncvoter & 1000 & 19 & 151 \\
abalone & 4177 & 9 & 191.9 \\
Students & 5000 & 21 & 516.81 \\
Biocase multimedia object & 18,784 & 15 & 9,100\\
Wine Reviews & 150,935 & 11 & 17,540 \\
Biocase gathering namedareas & 137,710 & 11 & 21,200\\
\bottomrule \end{tabular} \end{table}

\paragraph{Configuration}

To replicate the experiments from the original CFDFinder, \textit{SupportIndependentStrategy} and \textit{ConstantExpansionStrategy} were chosen as the main study strategies. The following values were selected as fixed pruning parameters: a minimum support gain of 5\% of the number of dataset rows, a maximum support drop of 10\%, a minimum confidence of 1.0, and a maximum pattern tableau length of 2000 patterns.

\subsection{Research questions}

We evaluated the performance and scalability of our proposed solution through a series of controlled experiments. This evaluation aimed to substantiate the advantages of implementing the algorithm in C++ with the proposed optimizations. The study addresses three primary research questions:

\begin{itemize}
    \item \textbf{RQ1}: What performance gains in terms of execution time and memory consumption are achieved by the optimized C++ version of CFDFinder compared to its Java implementation in Metanome?
    \item \textbf{RQ2}: To what degree does the multi-threaded ParCFDFinder surpass both the baseline implementation in Metanome and the single-threaded version of ParCFDFinder, and what overall speedup is achieved across a variety of datasets?
    \item \textbf{RQ3}: How effectively does ParCFDFinder scale as the number of execution threads increases from 1 to the limit of available physical cores, and how close is the achieved speedup to the ideal linear scaling?
\end{itemize}

\subsection{Experiments}

\textbf{Experiment 1.} In this initial experiment, we reproduce the evaluation scenario performed by the CFDFinder authors on the presented datasets in order to compare our proposed implementation options~--- the multi-threaded ParCFDFinder and the single-threaded ParCFDFinder~--- with the original Java baseline implementation.

\textbf{Experiment 2.} 
In the second experiment, we evaluate the scalability of three algorithm versions (the Metanome, single-threaded, and multi-threaded implementations of ParCFDFinder) with respect to two key dimensions: the number of rows and the number of columns.

\textbf{Experiment 3.} 
In the last experiment, we examine the scalability of ParCFDFinder under varying degrees of parallelism. By adjusting the number of threads in the pool, we measured the resulting impact on both execution time and memory consumption. 

\subsection{Discussion}

\textbf{Experiment 1.} 

\begin{table*}[htbp]
\small
\centering
\caption{Overall Results}
\label{tab:comparison}
\begin{adjustbox}{max width=0.9\textwidth}
\begin{tabular}{llrrrrr}
\toprule
Dataset & Algorithm & Time (s) & Mem. (MB) & \#CFDs & Speedup & Mem. factor \\
\midrule
\multirow{3}{*}{Avocado Prices$^\dagger$} & Metanome & TL & 3215.3 & 30 & -- & -- \\
 & ParCFDFinder (1 thread) & 727.35 & \textbf{5379.5} & 5285 & $>4.9\times$ & -- \\
 & \textbf{ParCFDFinder (8 threads)} & \textbf{157.67} & 5834.4 & 5285 & $> \textbf{22.8}\times$ & -- \\
\midrule
\multirow{3}{*}{BMW Global Sales} & Metanome & 463.14 & 2806.8 & 6468 & -- & -- \\
 & ParCFDFinder (1 thread) & 21.96 & \textbf{241.3} & 6468 & 21.1$\times$ & \textbf{11.6}$\times$ \\
 & \textbf{ParCFDFinder (8 threads)} & \textbf{6.59} & 270.9 & 6468 & $\textbf{70.3}\times$ & 10.4$\times$ \\
\midrule
\multirow{3}{*}{Biocase gathering namedareas$^\dagger$} & Metanome & TL & 3333.8 & 1 & -- & -- \\
 & ParCFDFinder (1 thread) & 46.22 & \textbf{948.8} & 80 & $>77.9\times$ & --  \\
 & \textbf{ParCFDFinder (8 threads)} & \textbf{30.11} & 2957.6 & 80 & $>\textbf{119.6}\times$ & -- \\
\midrule
\multirow{3}{*}{Biocase multimedia object$^\dagger$} & Metanome & TL & 6135.5 & 56 & -- & -- \\
 & ParCFDFinder (1 thread) & 72.63 & \textbf{769.2} & 256 & $>49.6\times$ & -- \\
 & \textbf{ParCFDFinder (8 threads)} & \textbf{19.44} & 3022.6 & 256 & $>\textbf{185.2}\times$ & -- \\
\midrule
\multirow{3}{*}{Bridges} & Metanome & 47.37 & 2723.9 & 13874 & -- & -- \\
 & ParCFDFinder (1 thread) & 9.83 & \textbf{141.8} & 13874 & 4.8$\times$ & \textbf{19.2}$\times$ \\
 & \textbf{ParCFDFinder (8 threads)} & \textbf{2.78} & 144.5 & 13874 & \textbf{17.0}$\times$ & 18.8$\times$ \\
\midrule
\multirow{3}{*}{Echocardiogram} & Metanome & 25.72 & 2814.5 & 9180 & -- & -- \\
 & ParCFDFinder (1 thread) & 3.66 & \textbf{123.5} & 9180 & 7.0$\times$ & \textbf{22.8}$\times$ \\
 & \textbf{ParCFDFinder (8 threads)} & \textbf{1.17} & 123.9 & 9180 & \textbf{22.1}$\times$ & 22.7$\times$ \\
\midrule
\multirow{3}{*}{Students$^\dagger$} & Metanome & TL & 2859.9 & 4717 & -- & -- \\
 & ParCFDFinder (1 thread) & 707.93 & \textbf{13897.2} & 84044 & $>5.1\times$ & -- \\
 & \textbf{ParCFDFinder (8 threads)} & \textbf{239.35} & 15153.3 & 84044 & $>\textbf{15.0}\times$ & -- \\
\midrule
\multirow{3}{*}{Wine Reviews$^\dagger$} & Metanome & TL & 3681.1 & 7 & -- & -- \\
 & ParCFDFinder (1 thread) & 310.62 & \textbf{4984.8} & 1439 & $>11.6\times$ & -- \\
 & \textbf{ParCFDFinder (8 threads)} & \textbf{102.20} & 5465.7 & 1439 & $>\textbf{35.2}\times$ & -- \\
\midrule
\multirow{3}{*}{Wisconsin Breast Cancer} & Metanome & 212.81 & 2823.2 & 4711 & -- & -- \\
 & ParCFDFinder (1 thread) & 18.25 & \textbf{160.0} & 4711 & 11.7$\times$ & \textbf{17.6}$\times$ \\
 & \textbf{ParCFDFinder (8 threads)} & \textbf{5.00} & 240.9 & 4711 & \textbf{42.6}$\times$ & 11.7$\times$ \\
\midrule
\multirow{3}{*}{abalone} & Metanome & 156.63 & 318.5 & 458 & -- & -- \\
 & ParCFDFinder (1 thread) & 1.81 & \textbf{130.6} & 458 & 86.5$\times$ & \textbf{2.4}$\times$ \\
 & \textbf{ParCFDFinder (8 threads)} & \textbf{0.66} & 153.8 & 458 & \textbf{237.3}$\times$ & 2.1$\times$ \\
\midrule
\multirow{3}{*}{ncvoter$^\dagger$} & Metanome & TL & 4090.9 & 4294 & -- & -- \\
 & ParCFDFinder (1 thread) & 7477.56 & \textbf{8012.9} & 326518 & $>2.9\times$ & -- \\
 & \textbf{ParCFDFinder (8 threads)} & \textbf{2153.15} & 8072.3 & 326518 & $>\textbf{10.0}\times$ & -- \\
\midrule
\multirow{3}{*}{nursery} & Metanome & 143.26 & 2799.2 & 255 & -- & -- \\
 & ParCFDFinder (1 thread) & 1.46 & \textbf{122.6} & 255 & 98.1$\times$ & \textbf{22.8}$\times$ \\
 & \textbf{ParCFDFinder (8 threads)} & \textbf{0.45} & 148.0 & 255 & \textbf{318.4}$\times$ & 18.9$\times$ \\
\bottomrule
\end{tabular}
\end{adjustbox}
\end{table*}

In this experiment, we evaluate the algorithm's performance on the datasets across several metrics: execution time, memory usage, and the number of mined CFDs. For the C++ implementations, we also report speedup and memory factor relative to the Java baseline. Execution time was limited to three hours; a dagger ($\dagger$) next to a dataset indicates that the Java version timed out on that test. The complete results are presented in Table~\ref{tab:comparison}.

On approximately half of the datasets, Java times out, while both C++ versions successfully complete all of them. When Java finishes within the time limit, its execution time and memory consumption are an order of magnitude higher than those of the C++ implementations. The multi-threaded version is on average 3--4$\times$ faster than the single-threaded one, though the speedup varies by dataset.

Memory usage of the multi-threaded version is nearly identical to that of the single-threaded one~--- an observation discussed in detail in Experiment~3. The memory improvement over Java stems from our optimizations and the absence of JVM overhead. For datasets where Java times out, memory factor is omitted.

Regarding the number of mined CFDs: when Java does not time out, our implementation mines the same number of CFDs (due to our methodology) but more rapidly. When Java times out, it mines an order of magnitude fewer CFDs than our C++ implementations.

It should be noted that in some cases where Java reached the time limit, our implementation utilized more memory. This is explained by the fact that Java managed to process only a negligible number of levels before reaching the time limit, which is confirmed by the small number of discovered CFDs.

\textbf{Experiment 2.}

\begin{figure}
    \centering
    \includegraphics[width=0.5\textwidth]{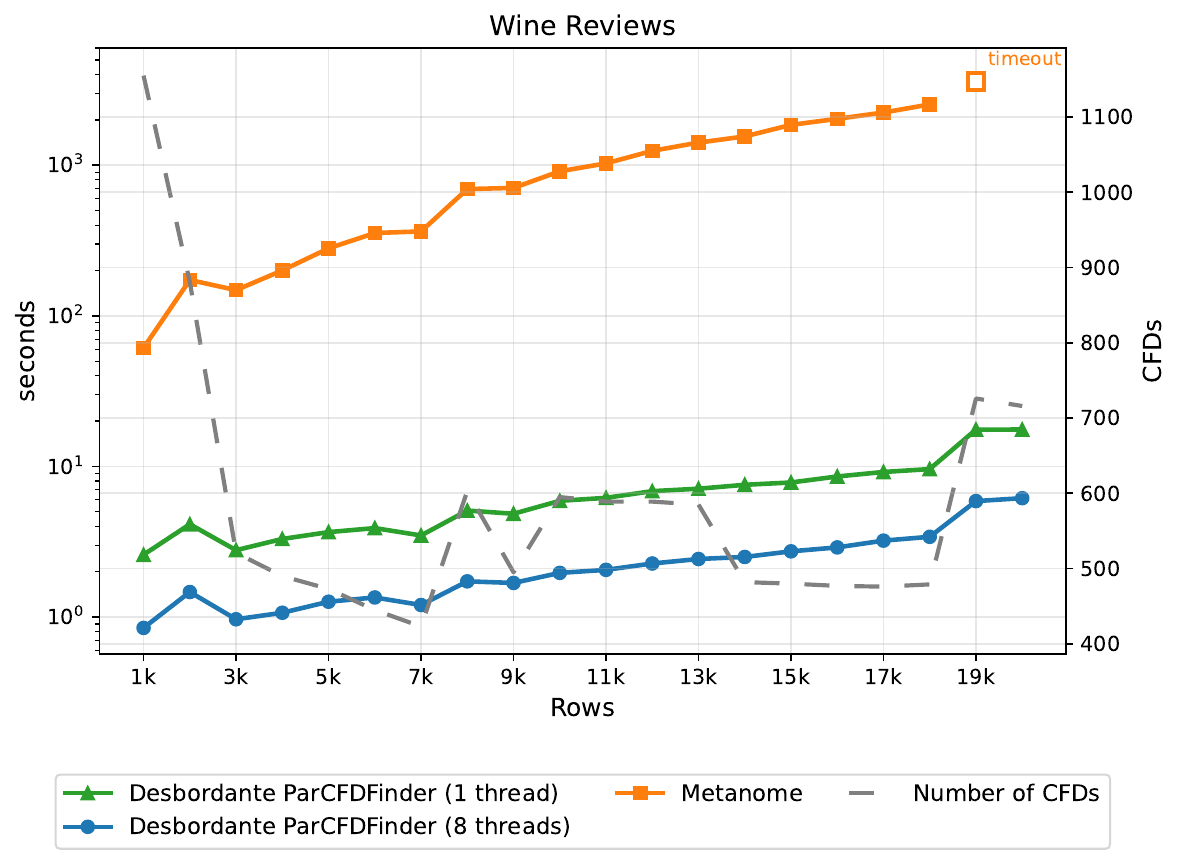}
    \caption{Scalability with respect to the number of rows: comparison of Java and C++ implementations on datasets up to 20,000 rows.}
    \label{fig:scalability_rows}
\end{figure}

\begin{figure}
    \small
    \centering
    \includegraphics[width=0.5\textwidth]{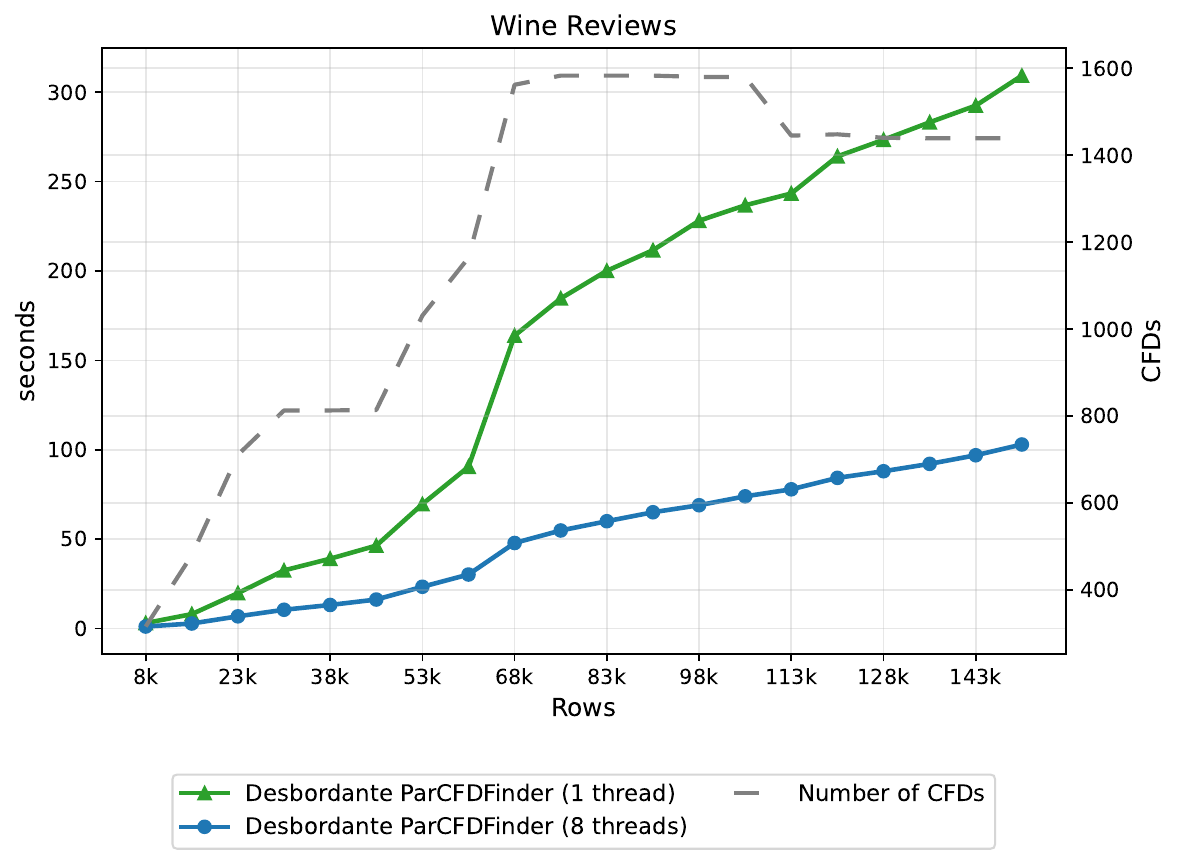}
    \caption{Scalability with respect to the number of rows: C++ implementations on the full dataset (150,935 rows).}
    \label{fig:scalability_rows_cpp_only}
\end{figure}

\begin{figure}
    \small
    \centering
    \includegraphics[width=0.5\textwidth]{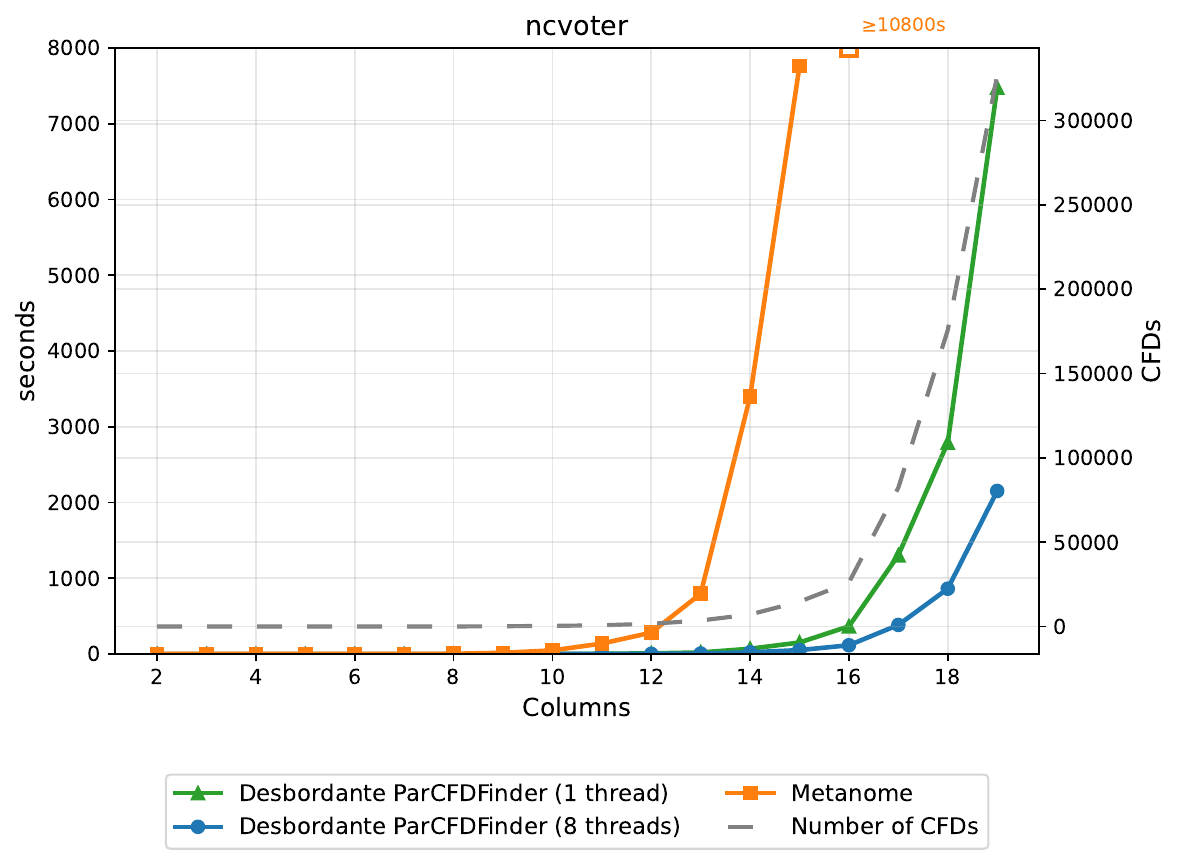}
    \caption{Scalability with respect to the number of columns.}
    \label{fig:scalability_columns}
\end{figure}

To evaluate row scalability, we use the \textit{Wine Reviews} dataset containing 150,935 rows. For column scalability, we use the \textit{ncvoter} dataset due to its large number of columns. These datasets were previously employed for scalability analysis in~\cite{cfdfinder}.

The results for row scalability are presented in two separate figures because the Java implementation fails to complete within the one-hour time limit once the number of rows exceeds approximately 18,000. In contrast, our C++ implementations efficiently process the full dataset within the time limit. Therefore, Figure~\ref{fig:scalability_rows} shows all three versions on smaller instances (where Java completes), while Figure~\ref{fig:scalability_rows_cpp_only} focuses on the C++ versions across the entire row range. For column scalability, both Java and C++ versions are presented in a single figure, as shown in Figure~\ref{fig:scalability_columns}.

Figure~\ref{fig:scalability_rows} presents the results on datasets truncated to 20,000 rows. The C++ implementations are several orders of magnitude faster than the Java version, and all three exhibit approximately linear growth in execution time as the number of rows increases. However, the Java implementation times out when the row count slightly exceeds 18,000, whereas both C++ versions process this scale successfully.

Figure~\ref{fig:scalability_rows_cpp_only} shows the benchmark on the full dataset. Both C++ implementations complete without timing out, and the execution time scales linearly with the number of rows~--- a favorable property for large-scale datasets. The multi-threaded version achieves a speedup of approximately 3$\times$ over the single-threaded C++ implementation on the largest instance.

Figure~\ref{fig:scalability_columns} demonstrates scalability with respect to the number of columns. All algorithms exhibit exponential growth in execution time, which is expected for a problem of this combinatorial nature. The Java version exhausts the three-hour time limit when the number of columns reaches approximately 15, while both C++ implementations successfully process all column configurations up to 19 columns. The multi-threaded version achieves a speedup of approximately 3--4$\times$ on the instance with the highest column count. As the number of columns increases, the advantages of parallelization become evident, and the multi-threaded version consistently outperforms the single-threaded one.

\textbf{Experiment 3.} 

\begin{figure}
    \small
    \centering
    \includegraphics[width=0.5\textwidth]{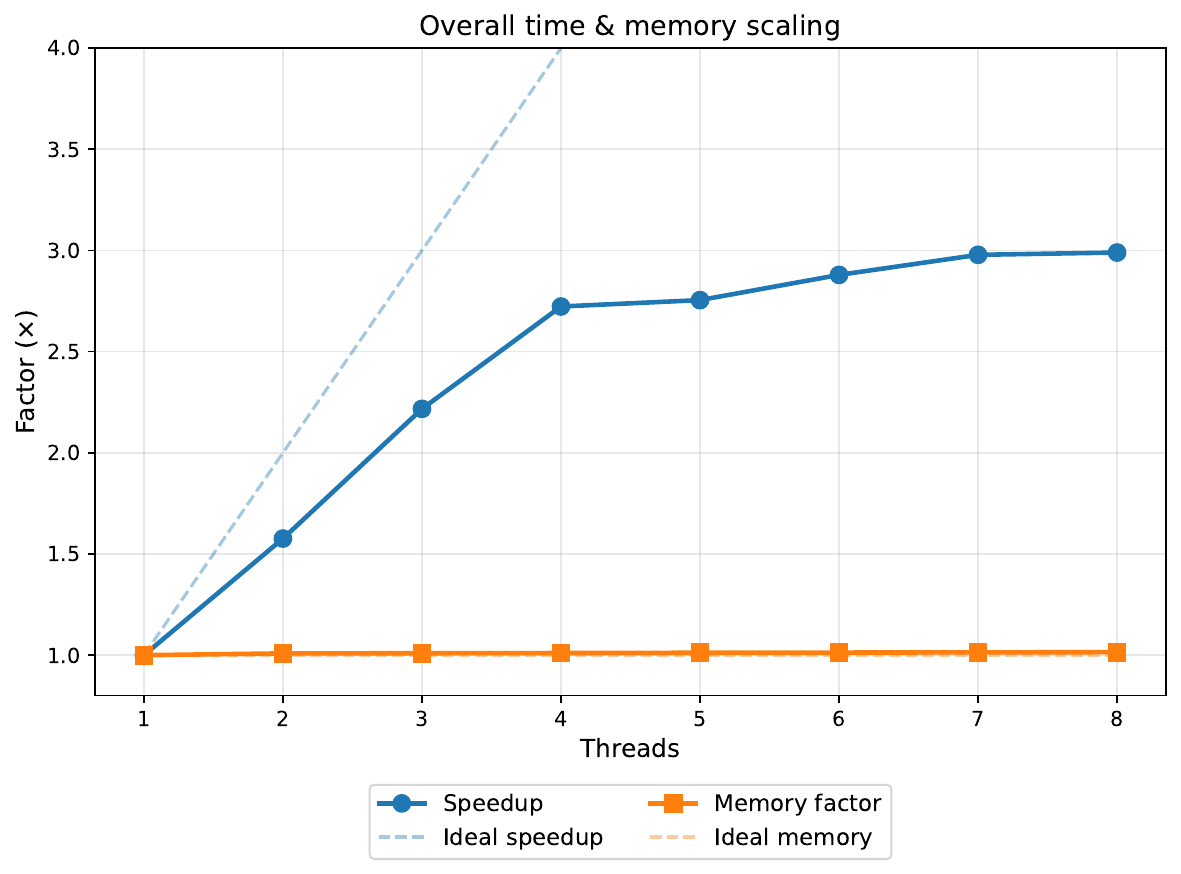}
    \caption{Scalability  with respect to the number of threads.}
    \label{fig:speedup}
\end{figure}

These measurements were conducted on the \textit{Students} dataset, as it demonstrated one of the highest runtimes and the largest number of detected CFDs across our datasets.

Figure~\ref{fig:speedup} shows the speedup achieved by the multi-threaded implementation as the number of threads increases, along with the corresponding memory consumption. We observe logarithmic growth in the speedup factor, which is a good result in practice. Additionally, the measured memory footprint remains very close to the baseline (single-threaded) level within the measurement variance.

Typically, increasing the thread count leads to higher memory consumption due to per-thread overhead (e.g., thread stacks and synchronization primitives). In our implementation, however, no such increase is observed. Two factors explain this behavior:

First, peak memory is dominated by the \textit{support\_map}, which stores CFD candidates with their support values for checking support decay during specialization. This structure accounts for 90--95\% of the total memory, making any per-thread overhead negligible.

Second, the algorithm traverses the pattern lattice from the top to the bottom level. The \textit{support\_map} accumulates candidates monotonically: all previous candidates remain stored, and new specialized ones are added at each level. Memory peaks at the deepest level, where the number of cumulative candidates is at its largest. Per-thread overhead at intermediate levels is effectively overshadowed by this monotonic growth.

Thus, any thread-related memory overhead is too small to be distinguished from the dominant \textit{support\_map} footprint, given our measurement precision.

\section{Case Study}\label{sec:case-study}

To evaluate the practical results of the algorithm, the implementation was tested on a dataset containing data regarding the effects and side effects of medications across various populations, titled ``1000 drugs and side effects''\footnote{URL:\url{https://www.kaggle.com/datasets/palakjain9/1000-drugs-and-side-effects/data}}. The dataset consists of 1,000 records and includes nine columns: \textit{Patient\_ID}, \textit{Age}, \textit{Gender}, \textit{Condition}, \textit{Drug\_Name}, \textit{Dosage\_mg}, \textit{Treatment\_Duration\_days}, \textit{Side\_Effects}, and \textit{Improvement\_Score}. 

The algorithm was executed with the same configuration used in the experiments, yielding a total of 269 CFDs. Below, we examine several of the discovered CFDs. 

First, consider the following CFD:
{\footnotesize
\begin{verbatim}
[Age,Condition,Drug_Name,Dosage_mg,
Treatment_Duration_days,Side_Effects] 
                -> Improvement_Score
PatternTableau {
        (_|Infection|_|_|_|_)
        (_|Diabetes|_|_|_|_)
        (_|Hypertension|_|_|_|_)
        (_|Depression|_|_|_|_)
}
Support: 0.792
Confidence: 1.0
\end{verbatim}
}
This CFD indicates that for conditions such as Infection, Diabetes, Hypertension, and Depression, if the patient's age, medication, dosage, side effects, and treatment duration are known, the treatment outcome (\textit{Improvement\_Score}) is uniquely determined. In other words, for these specific conditions, a particular therapy within a given age group yields a stable and reproducible result. The dependency has a support value of 0.792, demonstrating that it models a widespread pattern within the dataset. 

As another example, consider the following CFD:

{\footnotesize
\begin{verbatim}
[Age,Gender,Condition,
Treatment_Duration_days] -> Side_Effects
PatternTableau {
        (_|_|Pain Relief|_)
        (_|Female|Diabetes|_)
}
Support: 0.31
Confidence: 1.0
\end{verbatim}
}

In this sample, we specifically observe that for female patients with diabetes, the side effects of the treatment are entirely determined by their age and treatment duration (as shown in the second pattern of the tableau). This CFD has a support of 0.31 and a confidence of 1.0, representing a precise rule that holds true for specific patient subgroups.

\section{Conclusion}\label{sec:conclusion}

We have presented ParCFDFinder, an improved implementation of CFDFinder obtained by combining algorithmic and engineering techniques available in C++, together with a parallelization strategy that yields additional speedup. The resulting algorithm is up to $318\times$ faster than the state-of-the-art baseline and reduces memory usage by up to $23\times$. ParCFDFinder is integrated into Desbordante, an open‑source data profiler with a Python interface, enabling users to import and invoke the algorithm from Python using only a few lines of code.

Consequently, we have extended the practical applicability of CFD discovery: for the first time, users can process hundreds of thousands of records on a commodity machine within a reasonable time. This enables analysis of meaningfully sized datasets, extraction of actionable insights, and development of data‑quality applications. Our case study illustrated, through a concrete example, how ParCFDFinder serves these purposes.

Future work includes expanding the pool of CFD discovery algorithms available in Desbordante, as different algorithms yield complementary sets of CFDs and expose different tunable parameters.

\section*{Acknowledgments}\label{sec:ack}

We would like to express our gratitude to Thorsten Papenbrock and Felix Naumann for providing access to the text of the master's thesis~\cite{cfdfinder}.

\balance

\bibliographystyle{IEEEtran}

\bibliography{FRUCTexample}



\end{document}